\documentclass[aps,prb,reprint,superscriptaddress,longbibliography,floatfix]{revtex4-2}
\usepackage[T1]{fontenc}
\usepackage{amsmath,amssymb,bm,graphicx}
\DeclareFontShape{OT1}{cmss}{m}{n}{<->cmss10}{}
\DeclareMathSizes{9}{9}{7}{5}
\usepackage[colorlinks=true,linkcolor=blue,citecolor=blue,urlcolor=blue]{hyperref}
\hypersetup{pdftitle={Magnon binding in a quantum Hall ferromagnet with a periodic electric field},
 pdfauthor={Manas Kaushik Das}}
\newcommand{\ii}{\mathrm i}
\newcommand{\Tr}{\operatorname{Tr}}

\begin{document}
\title{Magnon binding in a quantum Hall ferromagnet with a periodic electric field}
\author{Manas Kaushik Das}
\email{manaskaushikdas@gmail.com}
\email{manaskaushik@iisc.ac.in}
\affiliation{Department of Mathematics, Indian Institute of Science, Bangalore 560012, India}
\date[]{}
\begin{abstract}
A spin wave in a filling-one quantum Hall ferromagnet is neutral
but carries an electric dipole perpendicular to its momentum.
A spatially periodic electrostatic potential therefore creates
parallel channels for spin waves. We use this established mechanism
to calculate how confinement changes the exchange binding of magnons
travelling in the same direction. Narrower channels increase the
overlap of their transverse wavefunctions and strengthen the
effective attraction. In an isolated harmonic channel, the energy
required to remove one magnon from a cluster of $N$ is
$\pi(k_*\ell)^4N(N-1)/4$ times the transverse level spacing to
leading order, where $k_*$ is the mean wave number along the channel
and $\ell$ is the magnetic length. We predict the frequency and
momentum dependence of the spectral line for adding a second magnon
to one already present, both in an isolated channel and in a
coupled array. Tunnelling in a wide array can make the binding
energy exponentially small, while a strip narrower than the
transverse pair size recovers one-dimensional binding at
sufficiently weak attraction. We also find that a moving potential
can carry a single wave packet with less spreading than free
propagation when its initial group velocity matches the potential's speed.
\end{abstract}
\maketitle
\raggedbottom
\section{Introduction}
How does confining a spin wave change its interaction with another
spin wave? We study a two-dimensional electron gas in a strong
perpendicular magnetic field. At filling one ($\nu=1$), one electron
occupies each orbital of the lowest Landau level, with all spins
aligned. Metal electrodes above and below the electron layer screen
the Coulomb interaction. We separately impose a smooth electrostatic
potential that varies periodically along $x$ and is uniform along
$y$. In a device, patterned electrodes (gates) could produce such a
modulation. Within the lowest Landau level the uniformly filled
spin branch is stationary under this potential. Its excitations
respond to the field even though the reference state stays fixed.

A spin flip replaces an up-spin electron by a down-spin electron,
leaving a positively charged hole in the up-spin branch. In a
magnon, the quantum of a spin wave, the centres of the electron and
hole cyclotron orbits are displaced perpendicular to the wave
momentum~\cite{kallin,cooper}. A magnon travelling along $y$ thus
carries a dipole along $x$. Its energy is lowest where the electric
field is strongest along the dipole. These minima confine the
magnon across $x$, leaving it free to propagate along $y$. Reversing
the momentum reverses the dipole and shifts the preferred channel
by half a period. Experiments in graphene have generated and
detected Hall magnons~\cite{wei} and sensed their dipoles through
their coupling to an interferometer~\cite{assouline}.

\begin{figure*}[!t]
\includegraphics[width=\textwidth]{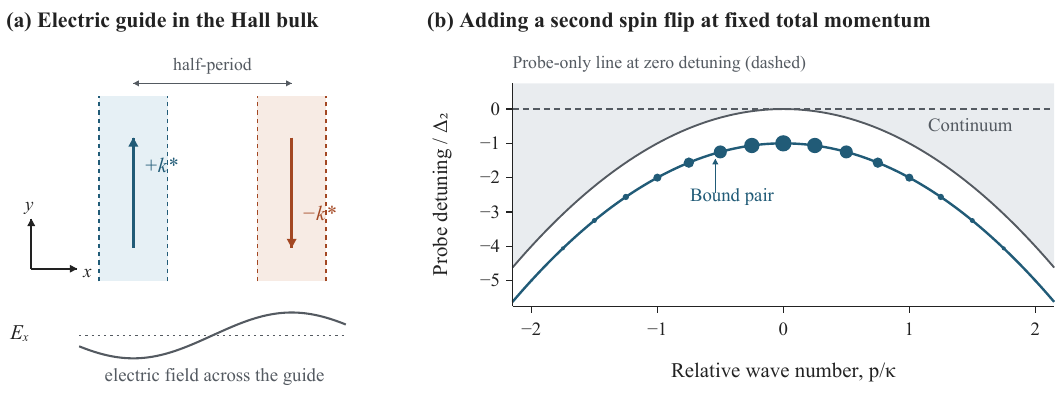}
\caption{Direction selection and pair spectroscopy.
(a) Magnons with opposite wave numbers along $y$ favour different
troughs. The dashed contours mark the confining regions across $x$;
arrows indicate the sign of momentum along $y$.
(b) The vertical axis shows the probe energy relative to that needed
to excite one magnon at the probe wave number $k_s$. The total wave
number of the two magnons is $2k_*$, with mean wave number $k_*$.
The shaded region shows the energies
available to two separated magnons. Marker areas show the calculated
transition strengths after normalization by the prepared population
and single-magnon excitation strengths. A larger circle means the probe is more likely to add a magnon to form the bound pair. The pair lies an energy
$\Delta_2$ below the continuum edge. The variation with relative wave
number $p$ reveals its inverse size $\kappa$.}
\label{fig:binding}
\end{figure*}

Bychkov and collaborators derived the equation for a single spin
exciton (the electron--hole description of a magnon) in a periodic
potential and found its confinement levels~\cite{bychkov,bychkov1997}.
Separate microscopic work established attraction and binding between
distinct Hall magnons~\cite{wojs,doretto}. We connect these two
results. Magnons sharing a spatial rotation of the spin direction
lower their exchange energy, and a narrower channel increases the
probability that they are close enough to interact. The electrostatic
potential can thus tune their effective attraction at fixed electron
interaction and filling.

We derive the effective interaction for magnons in the transverse
ground state and check its coefficient against microscopic electronic
states. The resulting one-dimensional Bose gas has a known exact
solution~\cite{mcguire,salasnich}. It gives a breakup energy in terms
of the measured spacing of the transverse levels and the magnon
momentum. For two magnons we calculate how the energy and strength
of the line for adding a second spin flip reveal the pair's binding
and size. We then follow the same line as tunnelling opens up
relative motion across neighbouring channels. The number of
channels determines whether that additional motion can be
treated as continuous at the binding energy.

We also study a translating potential. By comparing a single
magnon in a moving channel with the same initial packet in free
motion, we find that confinement can preserve the profile and
phase even when the freely propagating packet travels almost
the same distance.

\section{How the electric potential acts on a spin wave}
\label{sec:model}
We restrict the electron dynamics to the two spin branches of the lowest Landau level (LLL),
with the up-spin level filled at filling $\nu=1$. The magnetic length
$\ell=\sqrt{\hbar/(e|B|)}$, with $e>0$, sets the size of a Landau orbital.
The electronic kinetic energy is constant within the LLL, so exchange
sets the momentum dependence of the spin-wave energy. A uniform Zeeman energy $E_Z$ adds
$E_Z$ per reversed spin. We omit this constant until discussing
transition frequencies.
We take a clean, zero-temperature bulk with the polarized state
prepared. The interaction and electrostatic potential conserve the spin
component along the field, $S_z$, and leave this state stationary
within the LLL. Separate spin-flip fields create magnons and probe
their spectrum.

With the screening electrodes each a distance $d$ from the
electron plane, the two-dimensional Fourier transform of the
repulsion is
\begin{align}
 V_d(Q)&=\frac{2\pi g\tanh(Qd)}{Q},\notag\\
 E_C&=g/\ell,\qquad \delta=d/\ell,\qquad J=E_C\delta.
 \label{screening}
\end{align}
Here $Q$ is a wave number, $g$ includes the dielectric constant,
and $V_d$ has units of energy times area. The scale $E_C$ is the
Coulomb energy at separation $\ell$. With nearby electrodes the
exchange energy is of order $J$. The numerical calculations use
the full screened interaction.

Relative to the filled up-spin state $|F\rangle$, any state with
one reversed spin has the form
\begin{equation}
 |f\rangle=\sum_{dh}f_{dh}c^\dagger_{d\downarrow}c_{h\uparrow}|F\rangle,
 \qquad \Tr f^\dagger f=1.
 \label{one-flip-state}
\end{equation}
The indices $d,h$ label LLL orbitals. A scalar orbital operator $a_0$
acts on this amplitude as $a_0f-fa_0$. The two terms give the
responses of the added electron and the missing electron,
respectively.

For the normalized spin-exciton state $|\bm k\rangle$, the familiar
dispersion~\cite{kallin} is
\begin{align}
 \varepsilon(\bm k)&=D(0)-D(\bm k),\notag\\
 D(\bm k)&=E_C\int_0^\infty du\,\tanh(\delta u)e^{-u^2/2}
                  J_0(|\bm k|\ell u).
 \label{dispersion}
\end{align}
Here $J_0$ is the Bessel function of order zero. The exchange energy
is zero at $\bm k=0$
and approaches $D(0)$ at large momentum. At small momentum it is
quadratic:
\begin{align}
 \varepsilon(\bm k)&=\alpha|\bm k|^2+O(|\bm k|^4),\notag\\
 \alpha&=\frac{E_C\ell^2}{4}\int_0^\infty du\,u^2
                    \tanh(\delta u)e^{-u^2/2}.
 \label{stiffness}
\end{align}
For $d\ll\ell$,
$\varepsilon(\bm k)=J[1-e^{-|\bm k|^2\ell^2/2}]+O(J\delta^2)$
and $\alpha=J\ell^2/2+O(J\ell^2\delta^2)$.

A Fourier component of the electron density, measured relative to
the filled state and projected to the LLL, shifts the magnon wave
number according to
\begin{align}
 \delta\rho_{\bm Q}|\bm k\rangle
 &=2\ii f_Q\sin[\ell^2(\bm Q\mathbin\times\bm k)/2]
                  |\bm k+\bm Q\rangle,\notag\\
 f_Q&=e^{-Q^2\ell^2/4}.
 \label{vertex}
\end{align}
Here $\bm Q\times\bm k=Q_xk_y-Q_yk_x$, and $f_Q$ accounts for
the finite size of a Landau orbital. The orbit-centre coordinates
satisfy $[R_x,R_y]=-\ii\ell^2$~\cite{gmp,bychkov}. As derived in
Appendix~\ref{app:reduction}, the sine results from the difference
between the phases of the electron and hole responses. For a
potential smooth on the magnetic length, the leading term gives
the electric dipole
\begin{equation}
 \bm p=e\ell^2\hat{\bm z}\mathbin\times\bm k.
 \label{charge}
\end{equation}
Increasing the wave number separates the constituents further.
A wave travelling along $y$ therefore has a dipole along $x$.

We model the imposed modulation by the electron potential energy
\begin{equation}
 V(x,t)=A\cos(qx-\Omega t),\qquad A>0.
 \label{potential}
\end{equation}
At fixed $k_y\ne0$, shifting the origin to a minimum of the
magnon potential at $t=0$ gives the exact projected Hamiltonian
\cite{bychkov1997}:
\begin{align}
 H(t)&=\varepsilon(-\ii\partial_x,k_y)-V_0\cos[q(x-vt)],\notag\\
 V_0&=2A f_q|\sin(q\ell^2k_y/2)|,\qquad v=\Omega/q.
 \label{realspace}
\end{align}
The magnon well is displaced by a quarter period from a minimum
of the applied potential because the dipole responds to the field,
which is proportional to the gradient. At $k_y=0$ the dipole
component along $x$ vanishes. Opposite signs of $k_y$ favour
alternating wells [Fig.~\ref{fig:binding}(a)]. We first study
binding for a stationary potential, $\Omega=0$. In
Sec.~\ref{sec:transport} the wells translate along $x$.

\section{Why magnons in the same channel attract}
\label{sec:channel}
\subsection{The exchange energy supplies the attraction}
The origin of the interaction is easiest to see by describing a
slowly varying spin direction. We denote its polar angle by $\theta$,
its azimuth by $\phi$, and its projection along the field by
$m=\cos\theta$. The fully polarized
state has spin density $\hbar s_0$, with
\begin{equation}
 s_0=\frac{1}{4\pi\ell^2},\qquad n_m=s_0(1-m).
 \label{magnon-density}
\end{equation}
The density $n_m$ counts the spin deficit in units of $\hbar$:
$N=\int n_m\,d^2r$ is the number of magnons. The long-wavelength
exchange energy is the usual rotation-invariant ferromagnetic energy
\cite{sondhi,hofmann,canright},
\begin{align}
 H_{\rm ex}&=\frac{\rho_s}{2}\int d^2r\,
 \left[\frac{(\nabla m)^2}{1-m^2}+(1-m^2)(\nabla\phi)^2\right],\notag\\
 \rho_s&=\alpha s_0.
 \label{sigma-exchange}
\end{align}
The stiffness $\rho_s$ measures the energy cost of changing the spin
direction from point to point. Its value is fixed here by the exact
one-magnon dispersion, Eq.~\eqref{stiffness}.

For a constant tilt with azimuth $\phi=k_*y$, the spin direction
rotates along the channel at the wave number $k_*$.
Substituting $m=1-n_m/s_0$ into Eq.~\eqref{sigma-exchange} gives
\begin{equation}
 \frac{H_{\rm ex}}{\mathcal A}
 =\alpha k_*^2 n_m-\frac{\alpha k_*^2}{2s_0}n_m^2,
 \label{twist-energy}
\end{equation}
where $\mathcal A$ is the area. The first term is the energy of
independent magnons. The second is negative. Geometrically, the phase
gradient costs an energy proportional to the squared transverse spin,
$\sin^2\theta=2n_m/s_0-(n_m/s_0)^2$. As more spins tilt,
each additional flip contributes less to $\sin^2\theta$.
Magnons sharing the same phase twist consequently cost less
exchange energy than separated magnons. The leading attraction
vanishes at $k_*=0$,
where a uniform spin rotation costs no exchange energy.

Appendix~\ref{app:binding} checks this coefficient in the
electronic Hilbert space. For two equal-momentum flips in $M$
orbitals, the normalized exchange expectation relative to two
independent flips is $-2\varepsilon(\bm k)/(M-1)$. Its long-wavelength
limit agrees with Eq.~\eqref{twist-energy}. This tests the
interaction coefficient at equal incoming and outgoing momenta.
The binding energy follows from the relative motion in the guide.

\subsection{Transverse confinement sets the coupling}
We now take a stationary potential and magnons with mean wave
number $k_*>0$ along $y$. For $q\ell\ll1$ and $k_*\ell\ll1$,
the potential amplitude, oscillator length, and transverse
level spacing are
\begin{gather}
 V_0=Aq\ell^2k_*,\qquad
 L=\left(\frac{2\alpha}{V_0q^2}\right)^{1/4},\qquad
 \hbar\omega_x=\frac{2\alpha}{L^2}.
 \label{channel-scales}
\end{gather}
These follow by expanding $-V_0\cos(qx)$ at a minimum and matching
its curvature to the kinetic energy $-\alpha\partial_x^2$.
The transverse ground mode is
$u(x)=(\pi L^2)^{-1/4}e^{-x^2/(2L^2)}$; its position standard
deviation is $L/\sqrt2$.

We separate the dilute magnon field into its transverse mode,
carrier wave, and longitudinal envelope,
\begin{equation}
 \psi(x,y)=u(x)e^{\ii k_*y}\Phi(y).
 \label{channel-field}
\end{equation}
The carrier $e^{\ii k_*y}$ accounts for the mean momentum and
$\Phi$ describes the slower envelope. The spin dynamics make
$|\psi|^2=n_m$ the magnon number density. For two magnons in
this mode, the probability density for an encounter at $x$ is
proportional to $|u(x)|^4$.
Integrating over $x$ gives the overlap $I_4=\int |u|^4dx$.
For the Gaussian, $I_4=1/(\sqrt{2\pi}L)$ and the Hamiltonian
for the motion along the guide is
\begin{align}
 H_{\rm ch}&=-\alpha\sum_{i=1}^N\partial_{y_i}^2
              +g_{\rm 1D}\sum_{i<j}\delta(y_i-y_j),\notag\\
 g_{\rm 1D}&=-\frac{\alpha k_*^2}{s_0\sqrt{2\pi}L}.
 \label{channel-model}
\end{align}
We subtract the common one-magnon energy and the drift term,
which is linear in the total momentum measured from $N\hbar k_*$. 
The coefficient $g_{\rm 1D}$ of the delta-function interaction has
units of energy times length.
Spin symmetry fixes the exchange factor $-\alpha k_*^2/s_0$
\cite{hofmann}, and the confining potential sets the overlap $I_4$.
Appendix~\ref{app:binding} derives the interaction and shows why
the electrostatic self-energy and exchange cost of transverse
density gradients are small in a broad screened guide.

\subsection{Range of the one-dimensional description}
The reduction requires the magnon wavelength, transverse width, and
modulation period to be well separated:
\begin{gather}
 k_*\ell\ll1,\qquad q\ll k_*,\qquad
 \ell,d\ll L\ll q^{-1},\qquad k_*L\gg1,\notag\\
 \eta=N(k_*\ell)^2\ll1,\qquad
 (k_*\ell)^2|\log(L/\ell)|\ll1.
 \label{hierarchy}
\end{gather}
These conditions make the spin texture smooth on the magnetic
length, the channel broad compared with the screening distance
and narrow compared with the period. The condition $k_*L\gg1$
makes corrections from transverse gradients small. The parameter
$\eta$ keeps the internal binding below the transverse excitation
energy. The logarithmic condition keeps two-dimensional scattering
corrections perturbative before projection onto the channel.
We choose $2A<D(0)$ to separate high-momentum one-magnon states
from the guide. For the single-trough spectrum we further require
tunnelling to be small on the binding scale. Section~\ref{sec:crossover}
treats the pair when that condition is relaxed.

These requirements can hold simultaneously for many magnons:
$1\ll N\ll(k_*\ell)^{-2}$. The inverse length governing the internal motion is of order
$\eta/L$, the binding energy per magnon is of order
$\eta^2\hbar\omega_x$, and the peak spin-deficit fraction is of order
$\eta^2/(k_*L)^2$. The cluster is therefore long compared with its
width, weak enough to occupy a single transverse mode, and dilute
on the electronic spin scale. Appendix~\ref{app:binding} gives an
explicit scaling limit and the corrections to this leading theory.

\section{Binding energies and a spectroscopic test}
\label{sec:spectrum}
\subsection{The internal cluster spectrum}
McGuire's solution of the attractive delta-function Bose model
binds the relative separations while leaving the centre of mass
free~\cite{mcguire,salasnich}. At fixed total momentum $N\hbar k_*$, the leading channel theory
gives the cluster energy relative to $N$ magnons each at the
one-magnon energy $E_1(k_*)$,
\begin{equation}
 E_N-NE_1(k_*)=-\frac{g_{\rm 1D}^2}{48\alpha}N(N^2-1).
 \label{cluster-energy}
\end{equation}
The relevant continuum threshold is found by letting fragments
separate while keeping their total momentum fixed. Splitting into
$n$ and $N-n$ costs $g_{\rm 1D}^2Nn(N-n)/(16\alpha)$.
The lowest threshold removes one magnon, giving
\begin{gather}
 \Delta_N=\frac{g_{\rm 1D}^2}{16\alpha}N(N-1),\notag\\
 \boxed{\frac{\Delta_N}{\hbar\omega_x}
       =\frac{\pi}{4}(k_*\ell)^4N(N-1)}.
 \label{binding-prediction}
\end{gather}
Appendix~\ref{app:cluster} checks the contact boundary condition
and derives the fragment thresholds.

The spin stiffness and the potential amplitude cancel from
Eq.~\eqref{binding-prediction}. The absolute breakup energy still
increases with $A$ at fixed $q$ and $k_*$: a deeper well narrows the
mode as $L\propto A^{-1/4}$ and raises both the transverse spacing
and $\Delta_N$ as $A^{1/2}$. In the regime of Eq.~\eqref{hierarchy},
a cluster can break up along the channel at an energy far below
that needed to excite transverse motion.

For two magnons, the relative coordinate $r=y_1-y_2$ makes the
mechanism more transparent:
\begin{gather}
 H_r=-2\alpha\partial_r^2+g_{\rm 1D}\delta(r),\qquad
 \varphi_B(r)=\sqrt\kappa e^{-\kappa|r|},\notag\\
 \kappa=\frac{|g_{\rm 1D}|}{4\alpha}
       =\sqrt{\frac{\pi}{2}}\frac{(k_*\ell)^2}{L},\qquad
 \Delta_2=2\alpha\kappa^2.
 \label{pair-state}
\end{gather}
The derivative jump at $r=0$ balances the attractive contact energy
against relative kinetic energy. A weak attraction in one dimension
therefore gives a binding energy quadratic in its strength. The pair
length $\kappa^{-1}$ is much larger than $L$, explaining why a
transversely confined pair can still be very weakly bound.

\subsection{Adding the second spin flip}
We consider a two-pulse measurement of the binding energy and pair
size. A spin-flip pulse (the pump) creates a dilute population of
magnons near $k_*$ in the lowest transverse mode. A weaker pulse
(the probe) adds a second magnon with nearly the same momentum.
The leading transition driven by each pulse changes $S_z$ by one
quantum. The electrostatic potential confines the magnons throughout
this sequence.

The total one-magnon energy including the Zeeman contribution is
$\mathcal E_1(k)=E_Z+E_1(k)$. At equal pump and probe momenta the
bound-pair addition line is
\begin{equation}
 \hbar\omega_{\rm probe}=\mathcal E_1(k_*)-\Delta_2.
 \label{pair-readout}
\end{equation}
The probe-only one-magnon line is at $\mathcal E_1(k_*)$.
The new line is redshifted because the second magnon gains binding
energy upon joining the first. Its pump-dependent weight is
proportional to the prepared one-magnon population.

To resolve the pair's internal motion we hold the total wave number
at $2k_*$ and vary the relative wave number $p$. The pump and probe
then supply $k_p=k_*-p$ and $k_s=k_*+p$, respectively. The pair's Fourier amplitude is
$\widetilde\varphi_B(p)=2\kappa^{3/2}/(\kappa^2+p^2)$.
After accounting for the pump population and the measured
single-magnon excitation strengths, the transition weight is
\begin{equation}
 \frac{\mathcal W(p)}{\mathcal W(0)}
       =[1+(p/\kappa)^2]^{-2}.
 \label{pair-weight}
\end{equation}
At nonzero $p$, the first magnon has energy $\mathcal E_1(k_p)$.
The bound line is at
$2\mathcal E_1(k_*)-\mathcal E_1(k_p)-\Delta_2$.
Measured relative to the probe-only line $\mathcal E_1(k_s)$,
the separated-magnon threshold is lower by $2\alpha p^2$ at
leading order. The bound line is another $\Delta_2$ below it.
Figure~\ref{fig:binding}(b) shows this distinction: the gap to the
continuum stays fixed as $p$ varies, while the spectral weight
falls on the scale $\kappa$. Together the two measurements test
$\Delta_2=2\alpha\kappa^2$. Appendix~\ref{app:cluster} gives the
absolute weight for normalized modes.

Nonlinear spectroscopy can distinguish bound magnons from
single flips, as proposed for an anisotropic magnet and demonstrated
with momentum resolution in a trapped-ion spin chain~\cite{sato,kranzl}.
The proposed transition requires a spatially varying spin-flip
field carrying the required momentum. A uniform rotation leaves the
state in the same total-spin multiplet. Beams of finite extent
broaden the momentum dependence in Eq.~\eqref{pair-weight}. Resolving the
line calls for energy broadening small compared with $\Delta_2$
and momentum resolution finer than $\kappa$.

\section{Binding in an array of coupled channels}
\label{sec:crossover}
The transverse level spacing determines whether each magnon stays
in its lowest mode. Even in that mode, a magnon can tunnel to a
neighbouring channel. When tunnelling competes with binding,
two magnons can separate across the channels as well as along them.

We retain the lowest transverse band and model it with a tunnelling
matrix element $t_\perp>0$ between adjacent channels. The band has
width $W=4t_\perp$. For well-separated channels the leading
interaction acts when both magnons occupy the same channel. With
$\alpha_y$ denoting half the curvature of the one-magnon dispersion
along the channel, the isolated pair has binding energy
$\Delta_0=g_{\rm 1D}^2/(8\alpha_y)$. The channels repeat at
spacing $a=2\pi/q$, so the pair has a total transverse Bloch
wave number $K_x$ defined modulo $q$. At $K_x=0$ and total
longitudinal wave number $2k_*$, the binding energy $\Delta_{\rm arr}$
in an infinite array satisfies
\begin{equation}
 1=\int_{-\pi}^{\pi}\frac{d\vartheta}{2\pi}\,
 \left[\frac{\Delta_{\rm arr}}{\Delta_0}
     +\frac{2W}{\Delta_0}\sin^2\frac{\vartheta}{2}\right]^{-1/2}.
 \label{array-gap}
\end{equation}
The angle $\vartheta$ labels the relative wave number across the
channels in units of the inverse period. The integral sums over
these extra states of relative motion. Appendix~\ref{app:cluster}
derives it from the amplitude for the two particles to return to
the same point in the same channel. The method follows the
Green-function treatment of atomic pairs in periodic potentials
\cite{wouters}. Tunnelling there restores three-dimensional
relative motion. In the Hall array it restores two-dimensional
relative motion.

Writing $w=2W/\Delta_0$, the solution tends to
$\Delta_0-W$ to first order at small $w$. At large $w$,
$\Delta_{\rm arr}/\Delta_0\sim 16w\,e^{-\pi\sqrt w}$ in the
nearest-neighbour contact model. The return amplitude diverges
as an inverse square root at a one-dimensional threshold. With
a continuum of transverse states it diverges logarithmically.
The logarithm still gives a bound state, at an exponentially
smaller separation from the continuum. In this regime the
fractional change in the gap is approximately $\pi\sqrt w$ times
the fractional change in the interaction strength at fixed
dispersion. A quantitative prediction therefore requires
$\pi\sqrt w\,|\delta g_{\rm 1D}/g_{\rm 1D}|\ll1$. The response
at arbitrary $w$ follows from the bound-state normalization
in Appendix~\ref{app:cluster}.

\begin{figure}[!t]
\includegraphics[width=\columnwidth]{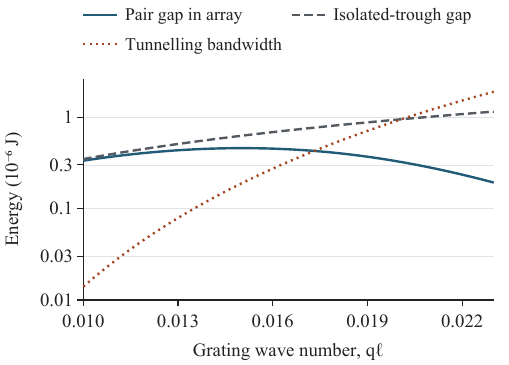}
\caption{Competing energy scales in a stationary periodic guide.
At $k_*\ell=0.20$, $A/J=0.10$, and $d/\ell=0.10$, shortening
the period increases both the overlap in one trough and the
tunnelling bandwidth $W$. The one-magnon band and transverse
mode are calculated with the full projected dispersion. The binding
curves use the leading contact interaction. The array curve uses the
first cosine harmonic of the lowest band, and $\Delta_0$ is the
binding energy in the absence of tunnelling.
The calculated gap to the second transverse band
exceeds $1.2\times10^{-4}J$.}
\label{fig:crossover}
\end{figure}

Figure~\ref{fig:crossover} shows why increasing transverse
confinement alone can be misleading. At fixed modulation amplitude and
mean momentum, decreasing the period narrows each trough and
increases the isolated-trough gap. But it also brings neighbouring
troughs closer together. In the illustrated sweep the array gap
peaks near $q\ell=0.015$ at about $4.6\times10^{-7}J$ before
falling, even though the gap to the second transverse band grows.
At three sampled settings we also use the calculated band beyond
its first harmonic in the binding equation. The correction to the
gap is about $0.03\%$ near the maximum and $1.5\%$ at the largest
$q\ell$ shown (Appendix~\ref{app:array-accuracy}). These comparisons
test the dispersion approximation within the contact model.

The infinite-array law also requires enough channels to resolve
the low transverse energies. For $N_c$ channels with periodic
transverse boundaries at $K_x=0$, the first nonzero transverse
energy of relative motion is $2W\sin^2(\pi/N_c)$. The integral
in Eq.~\eqref{array-gap} applies when this spacing is small
compared with $\Delta_{\rm arr}$. At fixed width
and sufficiently weak attraction only the lowest transverse
mode contributes to the threshold divergence. The pair then
binds as in one dimension, with the interaction reduced by the
probability that both magnons occupy the same channel.
Appendix~\ref{app:strip} gives the binding equation for a strip
with open boundaries.

The two-pulse measurement of Sec.~\ref{sec:spectrum} also
probes this change in relative motion. We take the pump and probe
wave numbers along the channels to be $k_*\mp p$ and the transverse
Bloch phases to be $K_xa/2\mp\vartheta$. In the nearest-neighbour
model their energy as separated magnons lies above the continuum
edge by $E_{\rm rel}=2\alpha_y p^2+2W_K\sin^2(\vartheta/2)$,
where $W_K=W\cos(K_xa/2)$ in the first zone. For this fixed
total momentum, the line for adding the second magnon and its
calibrated weight obey
\begin{align}
 \hbar\omega_s-\mathcal E_1(k_s,K_{xs})&=-\Delta_{\rm arr}-E_{\rm rel},
 \notag\\
 \frac{\mathcal W(p,\vartheta)}{\mathcal W(0,0)}
 &=\left(\frac{\Delta_{\rm arr}}{\Delta_{\rm arr}+E_{\rm rel}}\right)^2 .
 \label{array-addition}
\end{align}
Here $k_s=k_*+p$, $K_{xs}=K_x/2+\vartheta/a$, and
$\mathcal E_1(k_s,K_{xs})$ is the energy of the probe magnon
in the absence of the pump, and $\Delta_{\rm arr}$ at nonzero
$K_x$ is given by Eq.~\eqref{array-gap} with $W$ replaced by $W_K$.
The gap fixes the energy offset below the continuum and also
sets the longitudinal momentum width $\sqrt{\Delta_{\rm arr}/(2\alpha_y)}$.
For a shallow pair the weight narrows across the channels as
well, on the angular scale $\sqrt{2\Delta_{\rm arr}/W_K}$.
A source confined to one channel excites a range of total
transverse momenta, so its line samples different continuum
edges and binding energies.

\section{Transport by a moving trough}
\label{sec:transport}
The same electric confinement can preserve a single wave packet
during motion across the channels. We translate the potential by
one period $a=2\pi/q$ in time $T$, at speed $v=a/T$. At fixed
$k_y\ne0$, we centre the initial packet at a wave number $k_0$
whose group velocity matches $v$ and whose dispersion has positive
curvature,
\begin{equation}
 \varepsilon'(k_0,k_y)=\hbar v,\qquad
 \beta=\varepsilon''(k_0,k_y)/(J\ell^2)>0.
 \label{matching}
\end{equation}
Primes denote derivatives with respect to $k_x$. The preparation
pulse supplies the packet's initial momentum.
In the translating frame the moving trough balances the quadratic
dispersion, just as in a moving harmonic trap~\cite{chen}.
For $z=q\ell$, the local oscillator spacing and its momentum
variance are
\begin{equation}
 \hbar\omega_0=z\sqrt{\beta JV_0},\qquad
 (\sigma_k\ell)^2=\hbar\omega_0/(2\beta J).
 \label{oscillator}
\end{equation}
The initial Gaussian $b$ has variance $\sigma_k^2$ in momentum
and is centred at the bottom of the trough. Appendix~\ref{app:source}
specifies the spatial profile of a weak spin-flip field that
prepares this packet to leading order.

We compare the final packet with the translated input, using the
squared overlap
\begin{equation}
 \mathcal F=|\langle e^{-\ii ak_x}b|\psi(T)\rangle|^2 .
 \label{translated-fidelity}
\end{equation}
The same comparison for free propagation measures how much
the packet would spread at the same initial momentum. At fixed
$c=\omega_0T>0$, taking $q\ell\to0$ gives
\begin{equation}
 \mathcal F\longrightarrow1,\qquad
 \mathcal F_{\rm free}\longrightarrow(1+c^2/4)^{-1/2}.
 \label{shape-limit}
\end{equation}
Appendix~\ref{app:finite} bounds the error made by the harmonic
approximation in this limit. At finite wavelength the centre of
spin deficit $X_s$ moves at
$\dot X_s=\langle\partial_{k_x}\varepsilon\rangle/\hbar$.
The largest absolute group velocity,
$v_{\max}=\sup_{k_x}|\partial_{k_x}\varepsilon|/\hbar$ at fixed
$k_y$, limits the transport speed even in a deep well.

\begin{figure}[htbp]
\includegraphics[width=\columnwidth]{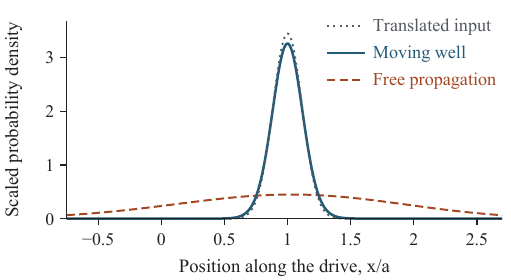}
\caption{Evolution of the same initial one-magnon packet over
the time $T$ for one period of translation ($JT/\hbar=400$).
The dotted curve is the input translated
by the target distance $a$. With the moving well the profile
stays close to this target, while free evolution spreads it.
The curves show probability density for the centre coordinate
along $x$, multiplied by $a$.
The physical spin profile also includes the electron--hole
separation and Landau-orbital smoothing.}
\label{fig:packet}
\end{figure}

For $q\ell=0.15$, $k_y\ell=0.4$, $d/\ell=0.1$, and $A/J=0.3$,
the full projected evolution at $JT/\hbar=400$ gives
$\mathcal F=0.99241$ and $\mathcal F_{\rm free}=0.25516$. The
driven and free centres move by $0.99967a$ and $0.98446a$,
respectively. The density profiles in Fig.~\ref{fig:packet}
show what the centres alone miss. This is a one-magnon
comparison. Resolving a bound pair requires the much longer
time set by its internal gap.

\section{Discussion and conclusions}
\label{sec:discussion}
An electrostatic potential can change the binding of spin waves
without changing the magnetic order through which they propagate.
This differs from using magnon transmission to probe magnetic
order across a junction between Hall states~\cite{weijunction}.
The single-magnon confinement mechanism of
Refs.~\cite{bychkov,bychkov1997} supplies the geometry. The
exchange attraction found in earlier studies of two Hall magnons
\cite{wojs,doretto} supplies the interaction. Here the geometry
determines how often the magnons sample that interaction. The
result is a prediction for the spectrum of two or more magnons
from the stiffness, the direction and magnitude of the carrier
momentum, and the transverse profile of a single magnon.

Transverse confinement is already a means of modifying scattering
in atomic gases~\cite{olshanii,salasnich}. In the Hall system,
the interaction itself depends on momentum: a uniform spin rotation
costs no exchange energy, and the attraction grows with the square
of the wave number of a shared spin twist. The same momentum
determines the electric dipole and hence the channel the magnon
occupies. The leading interaction and the curvature of the
one-magnon band along an isolated channel give
\begin{equation}
 \frac{\kappa_0}{I_4}=\pi\frac{\alpha}{\alpha_y}(k_*\ell)^2,
 \qquad \kappa_0=\frac{|g_{\rm 1D}|}{4\alpha_y}.
 \label{overlap-test}
\end{equation}
Here $I_4=\int |u(x)|^4\,dx$. The harmonic long-wavelength result
uses $\alpha_y=\alpha$ at leading order.
Changing the potential shape at fixed $k_*$ varies the overlap
$I_4$. Comparing channels at different $k_*$ after accounting for
that overlap tests the exchange dependence. The binding energy
and the momentum dependence of the pair line provide independent
estimates of $\kappa_0$ once the one-magnon dispersion is known.
Agreement would test the relative wavefunction as well as the
location of its energy level.

A periodic potential offers a second comparison. Near a harmonic
minimum at fixed $k_*$, the curvature of the magnon potential is
$V_0q^2\simeq A\ell^2 k_*q^3$. Varying $q$ and $A$ at fixed
$Aq^3$ would therefore keep the width of a single channel and
its predicted isolated binding energy fixed to leading order.
The distance between channels would still change, and with it the
tunnelling bandwidth. Within the smooth, deep-well regime and
the allowed range of amplitudes, such a comparison could isolate
the reduction of binding due to motion between channels.
A measurement at fixed total transverse wave number could compare
the pair line with the continuum edge calculated from the
measured one-magnon bands. An independently determined
isolated-channel binding energy $\Delta_{\rm ref}$ would fix
an effective coupling $|g_{\rm eff}|=\sqrt{8\alpha_y\Delta_{\rm ref}}$.
If the local mode and this coupling stay fixed as the channel
spacing varies, the measured band and the calibrated coupling
predict the tunnelling dependence of the line. The
isolated-channel data can then test Eq.~\eqref{overlap-test}
separately. For a finite strip the comparison also depends on
the transverse boundaries.

Electrical absorption and dipole-sensitive detection provide
ways to observe Hall magnons~\cite{noise,assouline}. Thermodynamic
measurements have also probed magnons bound to charged
textures~\cite{pierce}. The line proposed here concerns two
neutral spin waves in the same Hall fluid. A test requires a
spin-flip field that selects a finite wave number and couples
to the lowest transverse mode. The width of the line and the
probe bandwidth must both be smaller than the binding energy.
The momentum spread along the channel must be smaller than the
inverse pair length. In an array the spread of total transverse
momentum also broadens the transition because the pair and the
initial magnon disperse. A packet moving at velocity $v_y$ travels
$v_y\hbar/\Delta$ in the characteristic time needed to resolve a
gap $\Delta$. Near the maximum in Fig.~\ref{fig:crossover}, the
model gives a longitudinal pair length of about $1.4\times10^3\ell$
and a transit distance of about $4.2\times10^5\ell$, using the
group velocity along the channel from the one-magnon band. A
stationary probe of a localized packet would need to overlap it
over that distance. At fixed geometry in the isolated harmonic
limit, the spin stiffness cancels from this distance because it
sets both the group velocity and the binding energy.

The calculation suggests two extensions. First, a microscopic
two-spin-flip scattering calculation that includes higher
transverse modes could establish how far the leading interaction
remains accurate as the channel narrows or $k_*$ increases.
It would need to resolve the small binding energy relative to
the actual separated-magnon threshold. Second, the separation
of the oscillator and binding times suggests a dynamical probe
of the bond. A change in the modulation amplitude between these
two time scales could leave the magnons in the lowest transverse
mode while projecting part of the pair onto states of separated
magnons. For a larger cluster, the predicted dependence of the
removal energy on $N$ offers a test of successive binding in
the dilute regime. Its observation would require preparation
and spectroscopy of states with controlled numbers of magnons.

The moving-potential calculation addresses a different use of
the same confinement. It shows that the profile of a single
packet can survive translation even when a free packet travels
almost as far. For binding, the relevant comparison is between
the internal binding time and the time to tunnel between
channels. The transverse mode fixes the encounter probability
in one channel. The bandwidth and the transverse extent
determine how the magnons can separate, and therefore how the
binding energy and the momentum dependence of the line change
even when each magnon remains in the lowest band.

\appendix
\section{Exact one-spin-flip reduction}
\label{app:reduction}
We use the electron--hole description to calculate the spin-wave
energy for the screened interaction and recover the known coupling
to an electrostatic potential~\cite{kallin,gmp,bychkov,bychkov1997}.
For $M$ LLL orbitals per spin and $M$ electrons,
the coefficient matrices in Eq.~\eqref{one-flip-state} have inner product
$\langle f|g\rangle=\Tr f^\dagger g$. The many-electron operator $\rho(a_0)$ sums the one-electron
orbital operator $a_0$ over both spin species. Its action relative
to the filled state is
\begin{equation}
 [\rho(a_0)-\Tr a_0]|f\rangle=|a_0f-fa_0\rangle.
\end{equation}
The trace subtracts the filled-state expectation. The operator
$\tau_{\bm Q}=e^{\ii\bm Q\cdot\bm R}$ acts on the guiding centre,
which is the centre of a Landau orbit. Its products obey
\begin{align}
 \tau_{\bm Q}\tau_{\bm k}
 &=e^{\ii\ell^2(\bm Q\times\bm k)/2}\tau_{\bm Q+\bm k},\notag\\
 [\tau_{\bm Q},\tau_{\bm k}]
 &=2\ii\sin[\ell^2(\bm Q\times\bm k)/2]\tau_{\bm Q+\bm k}.
 \label{translation-algebra}
\end{align}
The form factor $f_Q$ supplies the remaining contribution from
the cyclotron orbital. Together these give Eq.~\eqref{vertex} for
the normalized spin exciton $|\bm k\rangle=|\tau_{\bm k}/\sqrt M\rangle$.

Relative to the filled state, the spin-independent interaction acts as
\begin{align}
 H_{\rm ex}f&=\frac{1}{2\mathcal A}\sum_{\bm Q}V_d(Q)f_Q^2
                     [\tau_{-\bm Q},[\tau_{\bm Q},f]],\notag\\
 \Tr f^\dagger H_{\rm ex}f
 &=\frac{1}{2\mathcal A}\sum_{\bm Q}V_d(Q)f_Q^2
                     \|[\tau_{\bm Q},f]\|_{\rm HS}^2.
 \label{exact-exchange-form}
\end{align}
Here $\mathcal A$ is the area, and $\|X\|_{\rm HS}^2=\Tr X^\dagger X$.
The constant due to normal ordering cancels against the filled-state
energy at fixed electron number. The matrices $\tau_{\bm k}$
diagonalize the double commutator. In the plane their energies are
\begin{equation}
 \varepsilon(\bm k)=\int\frac{d^2Q}{(2\pi)^2}V_d(Q)e^{-Q^2\ell^2/2}
                         [1-\cos(\ell^2\bm Q\times\bm k)].
\end{equation}
Angular integration gives Eq.~\eqref{dispersion}. Expansion at small
$\bm k$ yields the curvature in Eq.~\eqref{stiffness}. 

Both constituents contribute to the spin deficit. Adding a
down-spin electron and removing an up-spin electron each lower
$S_z$ by $\hbar/2$. With electron and hole number profiles $\rho_e$
and $\rho_h$, each normalized to one, we obtain
\begin{equation}
 w_s=(\rho_e+\rho_h)/2,\qquad
 \delta S_z=-\hbar w_s,\qquad X_s=\int xw_s\,d^2r.
\end{equation}
In the momentum representation $X_s=\ii\partial_{k_x}$.
The electron guiding centre minus the hole guiding centre is
$\ell^2(k_y,-k_x)$, which gives the dipole in Eq.~\eqref{charge}.
The electric term in Eq.~\eqref{realspace} commutes with $X_s$.
The speed bound follows from the largest absolute group velocity
in the exchange dispersion.

For screening electrodes close to the electron plane, the
approximation to the exchange kernel follows from
$0\le z-\tanh z\le z^3/3$, which gives
\begin{equation}
 |D(\bm k)/J-e^{-|\bm k|^2\ell^2/2}|\le2\delta^2/3.
\end{equation}
The finite-wavelength propagation uses the full integral throughout.

\section{Microscopic interaction and reduction to one dimension}
\label{app:binding}
\subsection{Exchange energy of equal-momentum spin flips}
We compute the interaction energy at equal magnon momenta while
keeping the electronic exclusion principle. The spin-lowering
operator $S^-_{\bm k}=c_\downarrow^\dagger Uc_\uparrow$ applies
the translation $U=\tau_{\bm k}$ during a flip, and
$S^+_{\bm k}=(S^-_{\bm k})^\dagger$. Unitarity of $U$ gives the
commutator and the normalized $n$-flip state below, with
$N_\downarrow$ the number of down-spin electrons.
\begin{align}
 [S^+_{\bm k},S^-_{\bm k}]&=M-2N_\downarrow,\notag\\
 |n;\bm k\rangle&=
 \frac{(S^-_{\bm k})^n|F\rangle}{\sqrt{n!M!/(M-n)!}}.
 \label{normalized-flips}
\end{align}
The factorial ratio counts the orbitals still available for flipping.

We evaluate the energy by forming a Slater determinant of the
$M$ orthonormal spinors
$\sqrt{1-p}|h,\uparrow\rangle+\sqrt p\,U|h,\downarrow\rangle$,
where $0\le p\le1$. The one-particle projector of this Slater
determinant is
\begin{equation}
 P_p=\begin{pmatrix}
 (1-p)I&\sqrt{p(1-p)}U^\dagger\\
 \sqrt{p(1-p)}U&pI
 \end{pmatrix}.
\end{equation}
For $A_Q=\operatorname{diag}(\tau_Q,\tau_Q)$, Wick contraction gives
the density fluctuation
\begin{align}
 &\Tr[P_pA_Q^\dagger(1-P_p)A_Q]\notag\\
 &\hspace{15pt}=2Mp(1-p)[1-\cos(\ell^2\bm Q\times\bm k)].
\end{align}
The mean total charge is uniform. Substitution into the projected
interaction, with the same filled-state subtraction as above, gives
\begin{equation}
 \langle H_{\rm ex}\rangle_p=Mp(1-p)\varepsilon(\bm k).
 \label{spiral-expectation}
\end{equation}
The probabilities of finding $n$ reversed spins are
$\binom Mn p^n(1-p)^{M-n}$. Since the interaction preserves their
number, its expectation is the binomial average of the energies
in the normalized states of Eq.~\eqref{normalized-flips}. The identity
\begin{equation}
 \sum_{n=0}^M\binom Mn p^n(1-p)^{M-n}
                  \frac{n(M-n)}{M-1}=Mp(1-p)
\end{equation}
and uniqueness of these polynomial coefficients give, for $M>1$,
\begin{equation}
 \langle n;\bm k|H_{\rm ex}|n;\bm k\rangle
 =\varepsilon(\bm k)\frac{n(M-n)}{M-1}.
\end{equation}
A finite-area formulation uses
the corresponding discrete one-magnon dispersion; the plane formula
is its thermodynamic limit.

Writing $n_m=Mp/\mathcal A$ and $M/\mathcal A=2s_0$ in
Eq.~\eqref{spiral-expectation} gives energy density
$n_m\varepsilon-\varepsilon n_m^2/(2s_0)$. The resulting interaction coefficient at equal incoming and
outgoing momenta is $-\varepsilon(\bm k)/s_0$, with leading term
$-\alpha|\bm k|^2/s_0$. This agrees with the long-wavelength
energy in Eq.~\eqref{twist-energy}, including its normalization.

\subsection{Magnon field and coupling to the electrostatic potential}
The spin Berry term, which supplies the time derivative in the
ferromagnetic action, makes the azimuth $\phi$ and spin deficit
$n_m$ a conjugate pair~\cite{sondhi,canright}. In our phase convention
the complex magnon field is $\psi=\sqrt{n_m}e^{\ii\phi}$, so that
$|\psi|^2$ counts magnons per unit area.
Using $m=1-n_m/s_0$, the exact density-gradient part of
Eq.~\eqref{sigma-exchange} is
\begin{equation}
 \frac{\rho_s}{4s_0n_m}
 \frac{(\nabla n_m)^2}{1-n_m/(2s_0)}.
\end{equation}
Its first term, together with the term linear in $n_m$ from the
phase energy, is $\alpha|\nabla\psi|^2$. Its next term is
$\rho_s(\nabla n_m)^2/(8s_0^2)$. The phase energy terminates at
quadratic order in $n_m$ and gives the negative term in
the following canonical energy:
\begin{align}
 H_{\rm ex}=\int d^2r\,\bigg[
 &\alpha|\nabla\psi|^2+
 \frac{\rho_s}{8s_0^2}(\nabla n_m)^2\notag\\
 &-\frac{\rho_s}{2s_0^2}n_m^2(\nabla\phi)^2+\cdots\bigg].
\end{align}
The dilute quantum Hamiltonian is normal ordered and its quadratic
coefficient is matched to Eq.~\eqref{stiffness}.

A smooth variation of the spin direction also changes the
electron number density. To leading order in gradients the
spin--charge relation~\cite{sondhi,canright} gives
\begin{equation}
 \delta n_e=\pm\ell^2\left[
 \partial_x(n_m\partial_y\phi)-\partial_y(n_m\partial_x\phi)\right].
 \label{texture-charge}
\end{equation}
The sign reflects the orientation of the magnetic field and selects
which minimum confines magnons at positive $k_*$. Since
$n_m\nabla\phi=\operatorname{Im}(\psi^*\nabla\psi)$, the coupling
of this charge to the prescribed potential is bilinear in $\psi$
at this gradient order. Integration by parts gives the dipole well
of Eq.~\eqref{channel-scales}. This one-magnon potential fixes
the transverse wavefunction $u(x)$ used below.

\subsection{Transverse overlap and repulsive corrections}
The transverse Gaussian in Eq.~\eqref{channel-field} gives
the overlap and density-gradient integrals
\begin{equation}
 I_4=\int |u|^4dx=\frac{1}{\sqrt{2\pi}L},\qquad
 \int[\partial_x(u^2)]^2dx=\frac{I_4}{L^2}.
\end{equation}
Keeping the carrier gradient $\partial_y\phi=k_*$ in the negative
phase term gives $g_{\rm 1D}=-\rho_sk_*^2I_4/s_0^2$, as in
Eq.~\eqref{channel-model}. The exchange energy associated with the transverse variation of
the magnon density adds
\begin{equation}
 \delta g_{\rm ex}=\frac{\rho_sI_4}{4s_0^2L^2}.
\end{equation}
Inserting the channel mode into Eq.~\eqref{texture-charge} gives
a charge density proportional to $k_*\ell^2\partial_x(u^2)|\Phi|^2$.
For $L\gg d,\ell$, the screened electrostatic self-energy of
this density variation is local on the scale of the mode. It adds
\begin{equation}
 \delta g_{\rm C}=V_d(0)k_*^2\ell^4\frac{I_4}{L^2}.
\end{equation}
Dividing by $|g_{\rm 1D}|$ gives $1/[4(k_*L)^2]$ for the exchange
gradient and $V_d(0)/(16\pi^2\rho_sL^2)$ for the Hartree term.
For $d\ll\ell$ the electrostatic correction relative to the
attraction is
$\ell^2/L^2+O(\delta^2\ell^2/L^2)$. Screening and a broad mode suppress both repulsive corrections.

\subsection{Scales governing the one-dimensional approximation}
The characteristic internal inverse length is
\begin{equation}
 |g_{\rm 1D}|N/(4\alpha)=O(\eta/L).
\end{equation}
Thus envelope momenta are
small compared with $k_*$, and the binding per particle and removal
gap are $O(\eta^2\hbar\omega_x)$. The corresponding peak fraction
$n_m/s_0=O(\eta^2/(k_*L)^2)$ is dilute. The binding energy per
particle compared with the transverse gap controls the deformation
of the occupied transverse mode. Virtual transverse excitations
also renormalize the two-body coupling. An intermediate state in a higher transverse mode costs energy
of order $\alpha/L^2$. Its longitudinal momentum integral gives
a relative correction of order $|g_{\rm 1D}|L/\alpha=O((k_*\ell)^2)$.
Summing over modes at shorter distances can introduce the
logarithm discussed below.

The lowest transverse energy, at leading harmonic order, gives the
one-magnon channel dispersion near positive $k$:
\begin{equation}
 E_1(k)\simeq\alpha k^2-Aq\ell^2 k+\frac{\hbar\omega_x(k)}2,
 \qquad \omega_x(k)\propto\sqrt k.
 \label{channel-dispersion}
\end{equation}
Consequently the coefficient of the quadratic energy for motion
along the channel is
\begin{equation}
 E_1''(k_*)/2=\alpha-\hbar\omega_x/(16k_*^2)
\end{equation}
at this order.
The relative correction to $\alpha$ is $1/[8(k_*L)^2]$.
The changes in $u(x;k)$ and $g_{\rm 1D}$ are also small over
the wave numbers spanned by the cluster.

Higher exchange gradients give relative $O((k_*\ell)^2)$ changes
to the leading interaction; trap anharmonicity gives $O((qL)^2)$.
Virtual scattering at distances between $\ell$ and $L$ still
samples two dimensions. For $g_{\rm 2D}=-\alpha k_*^2/s_0$, the
resulting relative correction to the interaction can be of order
\begin{equation}
 \frac{|g_{\rm 2D}|}{4\pi\alpha}\log(L/\ell)
 =(k_*\ell)^2\log(L/\ell).
\end{equation}
This is the origin of the logarithmic condition in
Eq.~\eqref{hierarchy}.

The scale inequalities can hold simultaneously. For example, at
fixed $A/J$ and $d/\ell$, the limit $\epsilon\to0$ with
$k_*\ell=\epsilon$, $q\ell=\epsilon^3$, and $N$ of order
$\epsilon^{-1}$ gives
\begin{align}
 L/\ell&\propto\epsilon^{-5/2},& k_*L&\propto\epsilon^{-3/2},\notag\\
 qL&\propto\epsilon^{1/2},&\eta&\propto\epsilon,\notag\\
 \hbar\omega_x&=O(J\epsilon^5),&\Delta_N&=O(J\epsilon^7).
\end{align}
Interwell tunnelling is exponentially smaller than these algebraic
scales. 

\section{Cluster energies and pair spectroscopy}
\label{app:cluster}
\subsection{An isolated trough}
The known internal McGuire wavefunction~\cite{mcguire} is
\begin{equation}
 \Psi_N\propto
 \exp\!\left[-\frac{|g_{\rm 1D}|}{4\alpha}
                  \sum_{i<j}|y_i-y_j|\right].
\end{equation}
Its logarithmic derivatives
in the ordering $y_1<\cdots<y_N$ are
$-|g_{\rm 1D}|(2j-N-1)/(4\alpha)$. Their squared sum in the
kinetic energy gives Eq.~\eqref{cluster-energy}. When two coordinates coincide,
the relative derivative jumps by $g_{\rm 1D}\Psi(0)/(2\alpha)$,
as required by the delta function. At fixed total momentum,
two separated clusters minimize their kinetic energy at equal
mean momentum per magnon. Inserting their populations $n$ and
$N-n$ into Eq.~\eqref{cluster-energy} gives a cost proportional
to $n(N-n)$. Further partition increases the energy because combining any two
fragments lowers their internal energy.

For the pair, $Y=(y_1+y_2)/2$ and $r=y_1-y_2$ separate the kinetic
energy into $-(\alpha/2)\partial_Y^2-2\alpha\partial_r^2$.
In a channel of length $\mathcal L\gg\kappa^{-1}$ the normalized
first-quantized bound wavefunction is
$\mathcal L^{-1/2}e^{\ii PY}\varphi_B(r)$, where $P$ denotes a
total wave number. Its relative Fourier transform is
\begin{equation}
 \widetilde\varphi_B(p)=\int dr\,e^{-\ii pr}\sqrt\kappa e^{-\kappa|r|}
                  =\frac{2\kappa^{3/2}}{\kappa^2+p^2}.
\end{equation}
The operator $a_k^\dagger$ creates a magnon in a normalized channel
mode. The matrix element for adding one magnon with the probe is
\begin{align}
 |\langle B,k_p+k_s|a_{k_s}^\dagger|k_p\rangle|^2
 &=\frac{2}{\mathcal L}|\widetilde\varphi_B(p)|^2\notag\\
 &=\frac{8\kappa^3}{\mathcal L(\kappa^2+p^2)^2},\qquad
 p=\frac{k_s-k_p}{2}.
 \label{absolute-pair-weight}
\end{align}
Symmetrization gives the factor two, including $k_s=k_p$, when
$a_k^\dagger|k\rangle$ has norm $\sqrt2$. The physical spin source
is matched to this canonical operator using the single-magnon orbital form factor and the dilute spin expansion. Dividing out those calibrated factors
and the prepared population gives Eq.~\eqref{pair-weight}.

At fixed local probe amplitude the coefficient of its normalized
mode grows as $\sqrt{\mathcal L}$, cancelling the factor
$\mathcal L^{-1}$ in the weight. For finite beams the matrix
element is averaged coherently over their Fourier envelopes.

\subsection{Binding energy and line strength in an infinite array}
We work at fixed total longitudinal wave number $2k_*$ and
transverse Bloch wave number $K_x$. In the nearest-neighbour
model, the sum of the transverse single-magnon energies depends on
the relative phase as $-W_K\cos\vartheta$ plus a constant, with
$W_K=W\cos(K_xa/2)\geq0$ in the first zone. Subtracting its
minimum and adding the longitudinal kinetic energy gives
$E_{\rm rel}$ of Sec.~\ref{sec:crossover}. This reduction uses a
constant longitudinal curvature and assumes that the binding and
the width of the lowest band are small compared with the gap to
the next band. The interaction acts at coincident longitudinal
positions in the same channel.

With $\lambda=|g_{\rm 1D}|$ and $\Delta=\Delta_{\rm arr}$, the
free propagator between coincident configurations below threshold
has matrix element
\begin{align}
 \Pi(\Delta,W_K)
 &=\left\langle\int\frac{dp}{2\pi}
       \frac{1}{\Delta+E_{\rm rel}}\right\rangle_\vartheta
 =\frac{\mathrm K(m)}
 {\pi\sqrt{2\alpha_y(\Delta+2W_K)}},\notag\\
 m&=\frac{2W_K}{\Delta+2W_K}.
 \label{array-return}
\end{align}
Here $\langle\cdot\rangle_\vartheta$ averages over one period with
measure $d\vartheta/(2\pi)$, and $\mathrm K$ is the complete
elliptic integral of the first kind. The contact interaction gives
the condition $\lambda\Pi=1$, which yields Eq.~\eqref{array-gap}
with $W_K$ in place of $W$. As $\Delta$ increases from zero to
infinity, $\Pi$ decreases from infinity to zero, giving a unique
bound state at fixed $K_x$. The asymptote
$\mathrm K(m)\sim\log(4/\sqrt{1-m})$ gives the exponential gap
of Sec.~\ref{sec:crossover}.

The bound-state equation gives a wavefunction proportional to
$(\Delta+E_{\rm rel})^{-1}$. Its squared norm before
normalization is the derivative of the return amplitude,
\begin{align}
 \mathcal N&=-\partial_\Delta\Pi
 =\frac{\mathrm E(m)}
 {2\pi\sqrt{2\alpha_y}\,\Delta\sqrt{\Delta+2W_K}},\notag\\
 \widetilde\varphi_B(p,\vartheta)
 &=\frac{1}{\sqrt{\mathcal N}\,(\Delta+E_{\rm rel})}.
 \label{array-wavefunction}
\end{align}
Here $\mathrm E$ is the complete elliptic integral of the second
kind. The wavefunction has unit norm in the measure
$dp/(2\pi)\,d\vartheta/(2\pi)$. At $W_K=0$ it is independent of
$\vartheta$ and becomes the isolated-channel transform with
$\kappa_0=\lambda/(4\alpha_y)$. For $\Delta\ll W_K$ its angular
scale is $\sqrt{2\Delta/W_K}$. The associated transverse decay
length is of order $a\sqrt{W_K/(2\Delta)}$.

In a large periodic normalization volume of $N_c$ channels
and length $\mathcal L$, the squared creation matrix element for
a normalized probe mode is
$2|\widetilde\varphi_B(p,\vartheta)|^2/(N_c\mathcal L)$.
The factor of two again comes from symmetrization. At fixed total
momentum the final energy is the separated-magnon threshold
minus $\Delta$. Subtracting the energy of the pump magnon gives
the probe transition energy. Subtracting the probe-only energy
as well gives the first line of Eq.~\eqref{array-addition}.
The normalized matrix element gives its second line.

The transverse part of the separated-magnon threshold is
$2\bar e-W_K$, where $\bar e$ is the centre of the one-magnon
band at $k_*$. The pair has transverse energy
$2\bar e-W_K-\Delta(W_K)$. In weak tunnelling, expansion of
Eq.~\eqref{array-return} gives
$\Delta(W_K)=\Delta_0-W_K+3W_K^2/(8\Delta_0)+\cdots$.
The terms linear in $W_K$ cancel in the pair energy. The
effective hopping of the pair between adjacent channels is
therefore $3t_\perp^2/(2\Delta_0)$ to leading order. For a
finite pump or probe, the observed line depends on the spread
of the energy difference between the pair and the initial
magnon across their momentum distributions.

\subsection{Accuracy of the array binding energy}
\label{app:array-accuracy}
Differentiating $\lambda\Pi=1$ with respect to the interaction
strength at fixed dispersion gives the fractional sensitivity
\begin{equation}
 S_g=\frac{\partial\log\Delta}{\partial\log\lambda}
 =\frac{1}{\lambda\Delta\mathcal N}
 =\frac{2\mathrm K(m)}{\mathrm E(m)} .
 \label{array-sensitivity}
\end{equation}
It is $2$ for isolated channels and grows to $\pi\sqrt w$ for
$w=2W_K/\Delta_0\gg1$. For $W_K>0$, rescaling the energies in
Eq.~\eqref{array-return} gives the first-order response to
simultaneous changes,
\begin{align}
 \delta\log\Delta={}&S_g\,\delta\log\lambda
 -\frac{S_g}{2}\,\delta\log\alpha_y \notag\\
 &+\left(1-\frac{S_g}{2}\right)\delta\log W_K .
 \label{array-variation}
\end{align}
A change in the potential can vary all three inputs. At fixed
$w$ the sensitivity stays finite in the asymptotic hierarchy
below. At large $w$ even a small correction to the interaction
can cause a substantial relative shift of the line.

In the smooth cosine potential, $b=qL$ measures the width of
a well in units of $q^{-1}$ and $\epsilon=k_*\ell$ measures
the longitudinal wave number in magnetic-length units. The
dimensionless tunnelling action at the well bottom is
$\int_0^{2\pi}d\xi\,\sqrt{2(1-\cos\xi)}/b^2=8/b^2$.
To logarithmic accuracy, $b^2\sim2/|\log\epsilon|$ makes
$W/(\hbar\omega_x)$ compete with the isolated pair scale
$\Delta_0/(\hbar\omega_x)\propto\epsilon^4$. Varying $q$ near this
scaling keeps $W/\Delta_0$ fixed as $\epsilon\to0$. At fixed $A$,
$q/k_*=A\ell^2b^4/(2\alpha)$ ensures $q\ll k_*$ and
$k_*L\gg1$. Thus the crossover can occur with both $W$ and
$\Delta_0$ small compared with $\hbar\omega_x$.

The same method accepts a measured or computed one-magnon band.
For transverse energy $e(\theta)$ at $k_*$, the relative
transverse energy is
\begin{align}
 \epsilon_K(\vartheta)&=s_K(\vartheta)-\min_{\theta}s_K(\theta),
 \notag\\
 s_K(\vartheta)&=e(K_xa/2+\vartheta)+e(K_xa/2-\vartheta).
\end{align}
The binding equation is then
$1=\sqrt{\Delta_0}\langle(\Delta+\epsilon_K)^{-1/2}\rangle_\vartheta$.
The normalization in Eq.~\eqref{array-wavefunction} follows
from the derivative of the same integral with its contact
prefactor, and the line calculation uses
$E_{\rm rel}=2\alpha_y p^2+\epsilon_K(\vartheta)$. In particular,
even Fourier harmonics of the band contribute at $K_xa=\pi$,
where the first harmonic cancels from the sum of energies.

For Fig.~\ref{fig:crossover} we diagonalize the stationary
one-magnon Hamiltonian in the basis of momenta separated by $q$,
using the full screened dispersion. The first cosine coefficient
of the lowest band gives $-2t_\perp$. Summing its Bloch states in
phase gives the wavefunction localized in one trough. We use its
overlap $I_4$ in the leading exchange interaction and use the
curvature at the bottom of the band for $\alpha_y$. Refining the
momentum cutoff and the numerical integrations changes the gaps
of this model by less than $5\times10^{-5}$ in relative value at
five sampled settings. At three settings we also retain eight
Fourier harmonics and solve the resulting integral at $K_x=0$.
The gap shifts by approximately $0.00047\%$, $0.032\%$, and
$1.50\%$ at $q\ell=0.010$, $0.01506$, and $0.023$, respectively.
The small band-shape correction therefore has a larger
fractional effect on the shallowest gap of these three points.
Both calculations use the same leading contact interaction.
The model estimate at $q\ell=0.01506$ gives $qL\simeq0.93$ in
the harmonic approximation, so the numerical mode is important
for this example. The power-counting parameter
$(k_*\ell)^2\log(L/\ell)$ is about $0.17$ there. Determining
the corresponding correction to the coupling requires
microscopic scattering information.

\subsection{Finite arrays and open strips}
\label{app:strip}
A finite periodic array has discrete relative phases
$\vartheta_j=2\pi j/N_c$ at $K_x=0$. Equation~\eqref{array-gap}
becomes
\begin{equation}
 1=\frac1{N_c}\sum_{j=0}^{N_c-1}
 \left[\frac{\Delta}{\Delta_0}
       +\frac{2W}{\Delta_0}\sin^2\frac{\pi j}{N_c}\right]^{-1/2}.
 \label{finite-array-gap}
\end{equation}
At fixed $N_c$ and sufficiently small $\Delta_0/W$, the $j=0$
term dominates and gives $\Delta\simeq\Delta_0/N_c^2$.
In this limit the magnons share a uniform transverse mode and
their probability of meeting in one channel is $1/N_c$.
The continuum of transverse phases is recovered when the lowest
nonzero transverse energy is small compared with the binding
energy. The required number of channels therefore grows as the
pair becomes shallower.

For open boundaries, let $\chi_m(j)$ be real orthonormal
single-magnon eigenmodes on the channels, with energies $e_m$
and a nondegenerate lowest level $e_0$. At fixed total
longitudinal wave number the continuum starts with both
magnons in this lowest mode. Resolving the propagator into
products of transverse eigenmodes gives its matrix elements
between coincident configurations in channels $i$ and $j$,
\begin{equation}
 \Pi^{\rm strip}_{ij}(\Delta)=
 \frac{1}{2\sqrt{2\alpha_y}}
 \sum_{m,n}
 \frac{\chi_m(i)\chi_n(i)\chi_m(j)\chi_n(j)}
 {\sqrt{\Delta+e_m+e_n-2e_0}} .
 \label{strip-return}
\end{equation}
The sum is over ordered pairs of modes, as in the product-space
resolution of the identity. The lowest bound energy is obtained
from $\lambda\,\lambda_{\max}(\Pi^{\rm strip})=1$, where
$\lambda_{\max}$ denotes the largest eigenvalue of this positive
matrix. The eigenvalue increases without bound as $\Delta\to0$
because both magnons can occupy $\chi_0$. In the weak-attraction
limit at fixed width this term gives
\begin{equation}
 \Delta_{\rm strip}\simeq\Delta_0
      \left(\sum_j|\chi_0(j)|^4\right)^2 .
 \label{strip-gap}
\end{equation}
The sum is the probability for two magnons in the transverse
ground mode to occupy the same channel. For a uniform open chain
with $N_c\geq2$ channels it equals $3/[2(N_c+1)]$ because the
ground mode is a half sine wave across the strip. At intermediate
coupling the other modes in Eq.~\eqref{strip-return} contribute.
For example, at $2W/\Delta_0=10$ and $N_c=16$, the gaps in units
of $\Delta_0$ are $0.0370$ for an open chain, $0.0218$ with
periodic boundaries, and $0.00774$ in the infinite array.

\section{The Gaussian transport limit and numerical evaluation}
\label{app:finite}
\subsection{Error of the harmonic approximation}
We bound the error accumulated over a translation by one period.
The dimensionless momentum and position are $p=k_x\ell$ and
$\xi=x/\ell$. In units of $J$ the dispersion is
$e(p)=\varepsilon(k_x,k_y)/J$ and the potential amplitude is
$w=V_0/J$. The dimensionless duration and packet width are
$\tau=JT/\hbar$ and $s_k=\sigma_k\ell$. The Hamiltonian in the
frame of the moving potential is
$e(p)-up-w\cos(z\xi)$, with $u=\hbar v/(J\ell)$.
Its quadratic expansion at $(p_0,0)$ has the matched Gaussian as an
exact eigenvector, and $s_k^2=z\sqrt{w/\beta}/2$.

Taylor's theorem bounds the kinetic remainder on this Gaussian by
$M_3\sqrt{15}s_k^3/6$, where $M_3=\sup_p|e'''(p)|$.
The Gaussian sixth moment supplies $\sqrt{15}$. The global cosine
inequality $|1-\cos r-r^2/2|\le r^4/24$ and the eighth position
moment bound the potential remainder by
$\sqrt{105}\,\beta z^2/96$. Duhamel's identity bounds the difference between exact and
harmonic evolution by the time integral of these remainders:
\begin{equation}
 \mathcal E\le\tau\left[
 \frac{\sqrt{15}}6M_3s_k^3+\frac{\sqrt{105}}{96}\beta z^2\right],
 \qquad \mathcal F\ge1-\mathcal E^2.
 \label{gaussian-residual}
\end{equation}
Here $\mathcal E$ bounds the norm of the difference between the
exact state and the translated packet after removing a common phase. Differentiating the
angular exchange integral and using $\tanh(\delta u)\le\delta u$
gives $M_3\le\sqrt{8/\pi}$.

At fixed $c=\omega_0T$, velocity matching gives
$u=2\pi\sqrt{\beta w}/c$. Positive curvature at $p=0$ provides
a small solution $p_0=O(z^{1/2})$. Since $w=O(z)$,
$T\propto z^{-3/2}$ and $s_k\propto z^{3/4}$, and
Eq.~\eqref{gaussian-residual} is $O(z^{1/2})$.
For free motion with quadratic dispersion, the squared overlap
with the translated input is
$[1+(\beta s_k^2\tau)^2]^{-1/2}$, with
$\beta s_k^2\tau=c/2$. The error from the quadratic approximation to the free dispersion
also vanishes, proving Eq.~\eqref{shape-limit}. 

\subsection{Exact finite-time formulas}
The periodic potential connects momenta $k_x=K+nq$ at each
fixed $K\in[-q/2,q/2]$~\cite{bychkov}. The operator $\mathsf n$
labels these momenta, $\mathsf n|n\rangle=n|n\rangle$, and the
shift $\mathsf T|n\rangle=|n+1\rangle$ increases $k_x$ by $q$.
The stationary Hamiltonian in this basis is
\begin{equation}
 H_0(K)=\operatorname{diag}_n\varepsilon(K+nq,k_y)
          -t_h(\mathsf T+\mathsf T^\dagger),\qquad t_h=V_0/2.
\end{equation}
Setting $\psi=e^{-\ii\Omega t\mathsf n}\chi$ gives
$\chi(t)=e^{-\ii H_R t/\hbar}b$, where the stationary operator
$H_R=H_0-\hbar\Omega\mathsf n$ is self-adjoint on the domain of
$\mathsf n$ because $H_0$ is bounded. For the numerical evolution we restrict the momentum index to a
finite interval. The resulting matrix has eigenvalues $E_j$ and
orthonormal eigenvectors $v_j$, with
$c_j=\langle v_j|b\rangle$, and
$J_{jl}=\langle v_j|\partial_KH_0|v_l\rangle$.
The integrated current at fixed $K$ is exactly
\begin{align}
 I_K=T\sum_{jl}c_j^*c_lJ_{jl}
 &e^{\ii(E_j-E_l)T/(2\hbar)}\notag\\[-2pt]
 &\times\operatorname{sinc}\!\left[\frac{(E_j-E_l)T}{2\hbar}\right],
 \label{finite-current}
\end{align}
where $\operatorname{sinc}x=\sin x/x$. With normalization
$q^{-1}\int_{-q/2}^{q/2}\|b(K)\|^2dK=1$,
\begin{align}
 \frac{\Delta X_s}{a}&=\frac{1}{2\pi\hbar}\int_{-q/2}^{q/2}I_K\,dK,\notag\\
 \mathcal F&=\left|\frac1q\int_{-q/2}^{q/2}dK\,
 e^{\ii Ka}b(K)^\dagger e^{-\ii H_RT/\hbar}b(K)\right|^2.
\end{align}
The integral adds the contributions from all $K$ sectors with
their weights and phases in the initial Gaussian. The free overlap follows by replacing $H_R$
by the diagonal laboratory-frame dispersion.

\subsection{Error from truncating the momentum basis}
The numerical basis ends at $n=\pm n_c$. When the numerical
state is extended by zero to the full basis, the couplings omitted
at these ends produce a residual of norm
\begin{equation}
 r_K(t)=t_h\sqrt{|\chi_{-n_c}(t)|^2+|\chi_{n_c}(t)|^2}.
\end{equation}
If the omitted initial norm is $e_K$, Duhamel bounds the full-space
state error by $e_K+\hbar^{-1}\int_0^t r_K(u)du$.
For a state of unit norm at fixed $K$, the resulting displacement
error is at most
\begin{equation}
 \frac{2v_{\max}}a\left[Te_K+
        \frac1\hbar\int_0^T(T-t)r_K(t)dt\right].
 \label{tail-current}
\end{equation}
For states of general norm at fixed $K$, we keep that norm in the
current estimate and integrate over $K$ with the normalization
above. Gaussian initial tails are
bounded by the complementary error function. The spectral sum also
gives a uniform-in-time envelope,
$|\chi_{\pm n_c}(t)|\le\sum_j|v_j(\pm n_c)c_j|$.
Thus the exterior estimate controls the integrated current as well
as the final state.

The screened integral is evaluated on $0\le u\le12$; the omitted
tail in $D/J$ is below $\delta^{-1}e^{-72}/12$.
Increasing the cutoff $n_c$ from $14$ to $18$, the $K$ quadrature
from $96$ to $192$ nodes, and the exchange quadrature from $640$
to $960$ nodes changes the reported displacements and fidelities
by less than $10^{-10}$. The error bounds due to the momentum cutoff are below
$4\times10^{-12}$. Independent laboratory-frame time integration
at three $K$ values agrees within $1.7\times10^{-11}$ in final
state norm and $2.1\times10^{-12}$ in the current integral in
units with $\ell=J=\hbar=1$.

\section{Spin-flip preparation and measurement}
\label{app:source}
A spin-flip field couples the filled state to the one-magnon sector.
Projection to the LLL multiplies its Fourier amplitude by
$e^{-|\bm k|^2\ell^2/4}$. We compensate for this factor to obtain
the desired Gaussian. The target amplitude $f$ and the required
Fourier amplitude $\widetilde\Omega$ of the field are
\begin{align}
 f(\bm k)&=h_y(k_y)c(k_y)
 e^{-(k_x-k_0)^2/(4\sigma_k^2)}e^{-\ii k_xx_0},\notag\\
 \widetilde\Omega(\bm k)&=e^{|\bm k|^2\ell^2/4}f(\bm k).
 \label{gaussian-source}
\end{align}
Here $h_y$ is a smooth, normalized envelope in $k_y$ with bounded
support away from $k_y=0$. The factor $c(k_y)$ normalizes the
$k_x$ profile, and $x_0$ is the initial well centre. The
$k_x$ dependence of the field remains Gaussian when
$\sigma_k\ell<1$. The support of $h_y$ in a bounded interval
keeps the inverse form factor bounded in $k_y$. The resulting
smooth field has finite integrated intensity. A Raman pulse can
provide a spin-flip operator; coherent control in a CdTe quantum
well in an in-plane field is one example~\cite{sweeney}. The
Fourier profile in Eq.~\eqref{gaussian-source} specifies the
field required in the present perpendicular-field geometry.

For a short weak pulse with normalized nonnegative temporal envelope
on $[-\tau_p,0]$ and carrier energy $E_s$, we hold the electric
potential stationary during the pulse, with one-magnon Hamiltonian
$H_0$. The first-order amplitude differs from its target
$b$ by at most
\begin{equation}
 \|b_p-b\|\le\frac{\tau_p}{\hbar}\|(H_0-E_s)b\|.
 \label{pulse-error}
\end{equation}
The bound follows by integrating
$\|(e^{\ii(H_0-E_s)t/\hbar}-I)b\|\le
|t|\|(H_0-E_s)b\|/\hbar$ over the pulse. This short-pulse estimate concerns preparation of the Gaussian
for transport. Resolving the pair line in a stationary channel
requires a longer probe pulse.

The first spatial moment of the spin density,
$P_z=\int x S_z(\bm r)\,d^2r$, measures displacement of the
spin deficit. This observable preserves the total spin component
along the field.
A weak spin-flip pulse creates the one-magnon state at first order
in its amplitude $\lambda$. The spin moment preserves the number
of flips, so its change during transport at fixed finite time is
\begin{equation}
 \Delta\langle P_z\rangle=-\lambda^2\hbar\Delta X_s+O(\lambda^4).
\end{equation}
Thus a spin-profile measurement averaged over repetitions gives
the displacement after calibration of the one-flip population.
Measuring the overlap in Eq.~\eqref{translated-fidelity} requires
a phase-sensitive comparison with the translated input.

\bibliography{pump-references}

\end{document}